# Ligand-Mediated Interactions Between Nanoscale Surfaces Depend Sensitively and Non-Linearly on Temperature, Facet Dimensions, and Ligand Coverage

*Asaph Widmer-Cooper*[*,1] *and Phillip L. Geissler*[2,3,4]

[1]School of Chemistry, University of Sydney, Sydney, NSW 2006, Australia, [2] Department of Chemistry, University of California Berkeley, Berkeley, California 94720, and [3]Materials Sciences Division and [4]Chemical Sciences Division, Lawrence Berkeley National Laboratory, Berkeley, California 94720

Corresponding author: asaph.widmer-cooper@sydney.edu.au.

**ABSTRACT**

Nanoparticles are often covered in ligand monolayers, which can undergo a temperature-dependent order-disorder transition that switches the particle-particle interaction from repulsive to attractive in solution. In this work we examine how changes in the ligand surface coverage and facet dimensions affect the ordering of ligands, the arrangement of nearby solvent molecules, and the interaction between ligand monolayers on different particles. In particular, we consider the case of strongly bound octadecyl ligands on the (100) facet of CdS in the presence of an explicit *n*-hexane solvent. Depending on the facet dimensions and surface coverage, we observe three distinct ordered states that differ in how the ligands are packed together, and which affect the thickness of the ligand shell and the structure of the ligand-solvent interface. The temperature dependence of the order-disorder transition also broadens and shifts to lower temperature in a non-linear manner as the nanoscale is approached from above. We find that ligands on nanoscale facets can behave very similarly to those on macroscopic surfaces in solution, and that some facet dimensions affect the ligand alignment more strongly than others. As the ligands order, the interaction between opposing monolayers becomes attractive, even well below full surface coverage. The strength of attraction per unit surface area is strongly affected by ligand coverage, but only weakly by facet width. Conversely, we find that bringing two monolayers together just above the order-disorder transition temperature can induce ordering and attraction.

Nanoparticles are appealing building blocks for creating new materials *via* bottom-up solution-phase processes.[1] Depending on the application, it can be desirable for the particles to be well dispersed or instead aggregated into structures with specific morphologies. Assessing and controlling the interaction between nanoparticles, a key factor in their assembly and stability to random aggregation or sintering, are therefore important challenges.

Inorganic nanoparticles in solution are often covered by ligand monolayers. These ligand shells not only influence the particles' growth[2] and physical properties,[3] but also strongly mediate interactions between them.[4] Recently, we have shown that octadecyl ligands on 4×20 nm CdS nanorods in *n*-hexane can undergo a temperature-dependent order-disorder transition in computer simulations, switching the rod-rod interaction from repulsive to attractive.[5] In this work we show that the temperature-dependence of this order-disorder transition – and thus the interaction between particles – is strongly affected by the density of ligand coverage and the scale of the various facet dimensions. Such sensitivities could be exploited in the laboratory to manipulate and optimize the assembly of ordered structures.

Macroscopic surfaces coated with surfactant monolayers have been widely studied, due in part to their industrial and technological importance.[6,7] Of the wide variety of systems that can form SAMs, alkanethiols on planar Au surfaces are the most widely studied.[7] Experimentally, it is known that such ligands can adopt structures that vary significantly with both surface coverage and temperature. For example, decanethiol ligands on Au(111) form stripes at low coverage (<27%), disordered SAMs at high coverage (>55%) and high temperature, and ordered SAMs at high coverage and low temperature.[8] In addition, the temperature dependence of the order-disorder transition on Au(111) is known to vary strongly with both surface coverage and chain length. For example, the transition is sharp at full coverage of decanethiol ligands but broadens and shifts to lower temperature when ligands are sparse, whereas the transition is broad even at full coverage for $SC_{22}H_{45}$ ligands.[8] How the ligands pack together and tilt relative to the substrate, has also been shown to depend on the chain length *n*, with distinct behaviors observed for $n \leq 14$ and $n \geq 16$.[9,10]

Force measurements between surfactant monolayers across hydrocarbon liquids were first performed more than 20 years ago.[11,12] These experiments show that the relationship between force and distance depends on both the degree of order of the monolayer as well as the shape of the solvent molecules. If the surface of the SAM is smooth and the solvent tends to form discrete layers near the surface (such as benzene or linear alkanes do), then the force-distance curve is generally oscillatory with a strong repulsion at short range as the ligands come into contact. In all other cases the force curve generally has a single broad minimum, with the hardness of the short-range repulsion depending mainly on the fluidity of the ligands. For sufficiently long ligands, such as end-grafted polymers, the interaction can even become purely repulsive.[13]

These extensive experimental measurements of SAM-mediated surface forces have been accompanied by relatively few computational studies. The repulsive interaction between oligo(ethylene glycol)-terminated alkanethiol SAMs in water has been examined,[14] but, to our knowledge, no simulations have addressed attractive forces between macroscopic SAMs in solution. As in experiments, simulations of linear alkanethiols on periodic Au(111) surfaces show that the ligands tilt strongly towards the substrate when they are ordered, both in vacuum (e.g. for $n = 16$)[15] and in hexane solvent (provided that $n > 6$).[16] The order-disorder transition was also found to be very broad for $n = 15\text{-}16$,[15,17] involving a gradual reduction in the tilt angle, followed by a rapid loss of tilt direction. More dramatically, studies of ligands bound to a generic triangular lattice have shown that the structure of the ordered state can vary strongly with the lattice spacing[18,19] and the chain length.[19] These observations suggest that a variety of distinct progressively ordered phases may exist, much as for liquid crystals.

While experimental characterization of ligand organization on nanoparticles in solution is extremely difficult and therefore rare, a large body of simulations and solid-state experiments point to key similarities with macroscopic SAMs. Experimental studies of dried Au nanoparticle films have noted that while dodecanethiol ligands on very small particles are poorly ordered under ambient conditions, the same ligands show spectroscopic properties similar to well-ordered SAMs on planar surfaces when the particles are larger than

~4.4 nm.[20] Longer octadecanethiol ligands have been observed to order on even smaller 3 nm Au particles[21], and simulations have shown that dodecanethiol ligands will align and form ordered bundles even on 1-2 nm Au particles if the temperature is sufficiently low.[22,23] In general, for small Au[24] and CdSe particles[25] the ordering temperature increases with chain length and radius of curvature. Changes in ligand structure with temperature have also been shown to affect the relative stability of both Au and PbS nanoparticle superlattices.[22,26,27]

On the other hand, it is still far from clear when ligand ordering on nanoparticles becomes an important effect in solution. For example, experimental studies indicate that under ambient conditions dodecanethiol ligands are disordered on 1-5 nm Au particles in carbon tetrachloride,[28] toluene and dichloromethane,[29] and molecular simulations are largely consistent with these observations.[30,31] In addition, simulations predict that 1 nm dodecanethiol-coated Au particles[31] and 5 nm PEO-coated Si particles[32] – which have disordered ligand shells near ambient conditions – will repel one another in *n*-hexane and water, respectively. By contrast, simulations have shown that ligands can align into ordered patches in poor solvents,[30,33] and our recent work indicates that octadecyl ligands on CdS particles with extended (~20 nm) facets in at least one dimension can order near room temperature even in a relatively good hexane solvent.[5] The well-characterized ordering behavior of macroscopic SAMs may therefore also be relevant to the interaction between, and assembly of, larger ligand-passivated nanoparticles in solution.

In this work we examine the ordering of strongly bound octadecyl ligands on Wurzite CdS(100) as a function of temperature and surface coverage in the presence of an explicit *n*-hexane solvent. In addition, we study the emergence of finite size effects as the two unique dimensions of the CdS(100) surface approach the nanoscale. To assess the impact of these changes on assembly, we determine potentials of mean force ($\phi_{MF}$) between opposing surfaces. Our calculations constitute the first study of monolayer ordering in solution as a function of temperature, surface coverage and facet dimensions – ranging from the planar macroscopic limit to the nanoscale – and of the variation in surface forces between the resulting SAMs.

**RESULTS**

**Macroscopic Surfaces.** We focus first on the ordering of ligands bound to periodically replicated CdS surfaces. This effectively macroscopic behavior provides an important reference for understanding finite size effects at the nanoscale. Our results echo those described by Hautman and Klein[15] for hexadecanethiol on Au(111), but differ in subtle ways that highlight the role of the substrate crystal structure and the ligand surface coverage.

To describe these differences it is useful to define several angles characterizing ligands' orientation relative to the substrate. A ligand's end-to-end vector is in general tilted away from the vector normal to the surface to which it is bound. This tilt has two distinct components (see Figure 1f) due to the anisotropic Wurzite lattice, one along the direction $x$ in which the Cd atoms are closest together (the angle $\theta_x$) and the other along the direction $z$ that is normal to $x$ and parallel to the surface (the angle $\theta_z$). Precise mathematical definitions of these tilt components are given in the SI (see Figure S2). The high- and low-temperature states of the ligands differ in the statistics of both of these angles.

For fully covered, periodic, planar surfaces, molecular dynamics simulation snapshots (e.g., Figures 1a and 1b) reveal a distinct change in ligand organization over a narrow range of temperature near 360 K. At higher temperatures, ligands sample a wide range of conformations and do not align closely with one another. Their orientations are, however, biased by the structure of the underlying CdS(100) surface. In particular, polarity of the crystal structure results in an average tilt of ligands towards the positive $z$ direction, even at very high temperature (390 K). Projection of ligands' end-to-end vectors onto the $x$ direction is by contrast distributed symmetrically around zero, reflecting the equivalence of $\pm x$ directions in the CdS lattice.

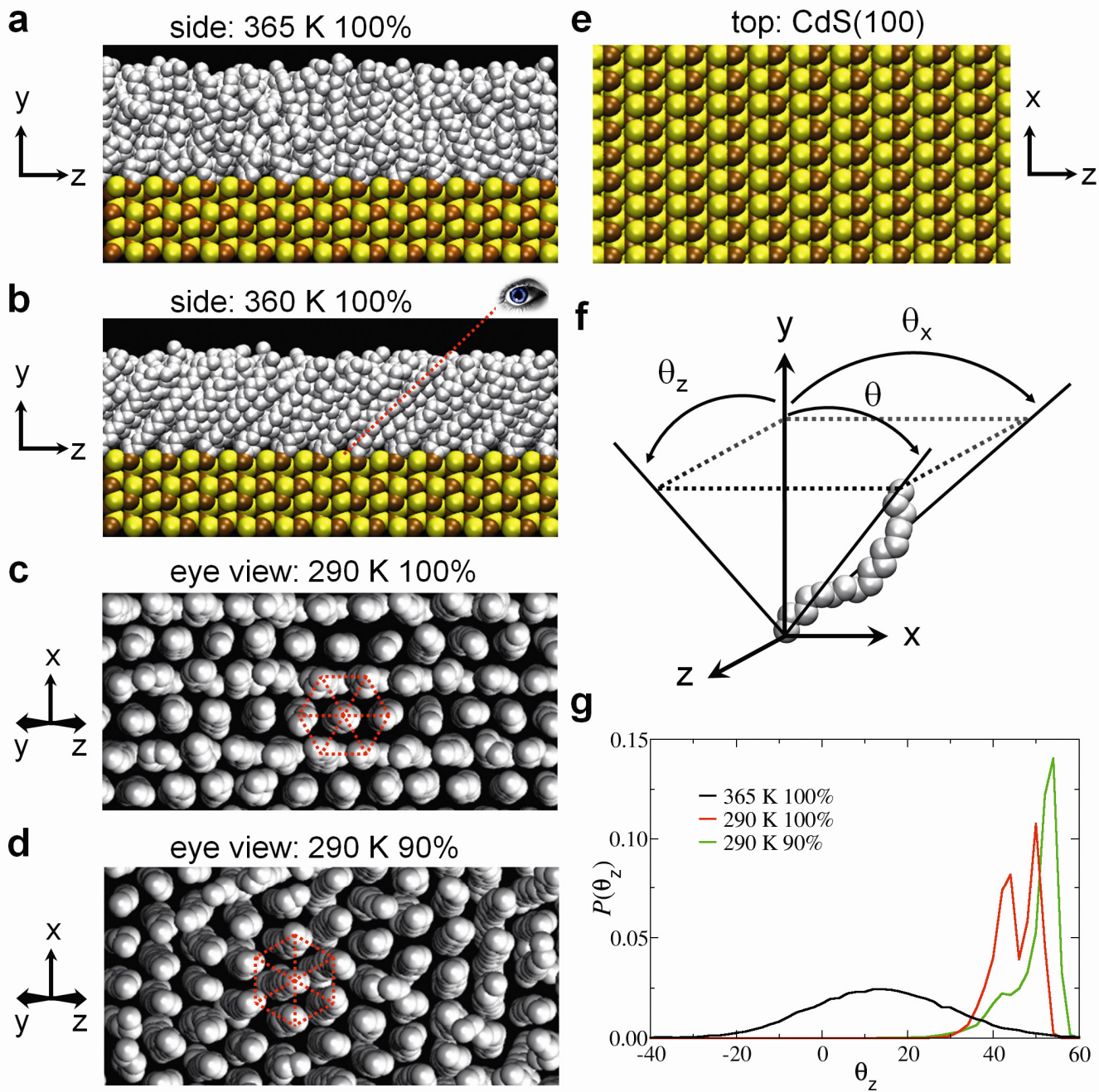

Figure 1. Ordering of octadecyl ligands (shown in white) on a periodic (100) surface of CdS (shown in brown and yellow, respectively) in *n*-hexane solvent (not shown here) at 100% and 90% passivation of the surface Cd atoms. Configurations (without solvent) are shown at (a) 365 K, (b) 360 K, and (c,d) 290 K. To emphasize the ligand packing in (c,d), the substrate is not shown and the ligands are viewed at an angle $\theta_z \approx 40°$. (e) Top view of the CdS(100) surface, (f) definition of ligand tilt angles, and (g) the ligand angle distribution $P(\theta_z)$ for selected systems, illustrating quantitative differences between the two ordered states in (c) and (d). Precise mathematical definitions of these tilt components are given in the caption of Figure S2.

At temperatures below 360 K, ligands adopt extended conformations, tilt more strongly away from the surface normal, align more closely with one another, and subtly break symmetries present in the higher temperature ('disordered') state. Viewed along the average tilt vector (see Figure 1c), the chains ends are packed together in a hexagonal pattern. This arrangement differs fundamentally from the rectangular array of Cd attachment sites on the CdS(100) surface. The two ends of a ligand molecule thus inhabit lattices with distinct symmetry. At 100% coverage, this lattice transformation is achieved by ligands tilting along the $z$ lattice vector, with those in every second row in Figure 1c adopting a *gauche* (rather than *trans*) dihedral angle near the CdS surface. This conformational ordering, in which ligands of each row are laterally displaced relative to those of adjacent rows, gives rise to two peaks in the $z$-angle distribution $P(\theta_z)$ (see the red line in Figure 1g). On Au(111), the crystal surface is itself hexagonally packed, and such subtle schemes are not required in order for the chains to pack densely; instead they simply tilt towards their next-nearest neighbors.[7] Aside from these details, the ordered state we have described on CdS(100) is similar to the Locked Rotator state described by Hautman and Klein[15] for hexadecanethiol on Au(111), i.e., ligands can rotate about their end-to-end vectors but the collective tilt direction is locked and almost all torsional angles are in the *trans* conformation.

In contrast, the loss of ligand order on CdS(100) as temperature increases proceeds quite differently from that observed for hexadecanethiol monolayers on Au(111). Hautman and Klein[15] observed a smooth and gradual decrease in the collective tilt angle ($\langle\theta\rangle$, see Figure 1f) from ~30° to 0° over a 300 K interval. For the CdS(100) surface, whose polarity inhibits collective rotation of the tilt direction, we instead observe a sharp change in the collective tilt angle (which is almost identical to the average $z$-tilt in our system, $\langle\theta\rangle \approx \langle\theta_z\rangle$). The resulting disordered state is characterized by both a small collective tilt angle and a high proportion of *gauche* torsional angles. This state is distinct from both of the high-temperature states described by Hautman and Klein. Together, these results demonstrate that changing the chemical identity of the surface underlying a ligand monolayer can affect both the detailed structure of the ligands' ordered state and the qualitative nature of the order-disorder transition.

**Effect of Reducing Ligand Coverage.** For the nanoscale semiconductor particles that we are interested in, the ligand coverage in experiments is typically neither well controlled nor easily characterized. The state of complete passivation on CdS(100) should be achievable, since the attachment site density on this surface (3.6/nm$^2$) is lower than that observed experimentally for both alkanethiol SAMs on Au[7] and octadecylphosphonic acid monolayers on water[34] (both ≈4.5/nm$^2$). Nonetheless, for less sticky ligands, and for nanoparticles that have been "washed" or aged, lower coverage states have been observed. Detailed characterisation of strongly-bound *n*-octylphosphonate ligands on 4-6 nm CdSe particles indicates greater than 85% surface coverage,[25] while studies of other ligands on CdS and CdSe particles report densities ranging from 0.6-4.6 ligands/nm$^2$.[35–38] Broad density ranges have also been reported for oleic acid ligands on PbS and PbSe particles,[39,40] which together with the CdS and CdSe results indicates greater variability of ligand coverage on semiconductor particles than has previously been observed for alkanethiols on Au nanoparticles[20] (where the density usually exceeds 4.5/nm$^2$). Greatly reduced ligand density clearly negates the ordering transition. Our results indicate that smaller changes in coverage can also substantially influence phase behavior, specifically the location and sharpness of the transition and even the structure of the ordered state.

As the fraction of Cd atoms that are passivated is reduced in simulations, the onset of ordering shifts to lower temperature and becomes less sharp. Defining $T_{order}$ as the maximum temperature at which $\langle \theta_z \rangle > 35°$ (see the SI for justification), we find that strong alignment persists up to $T_{order} \approx 345$ K at 90% coverage (see Figure 2). This crossover temperature decreases to $T_{order} \approx 312$ K at 75% coverage, accompanied by enhanced tilt angle fluctuations (see Figure S2) and broadening of the transition. At 65% coverage the ligands only order only weakly, and at even lower temperature ($\langle \theta_z \rangle \approx 25°$ at 290 K). We observe no significant ligand ordering at 50% coverage.

In these cases of incomplete passivation, the ordered states adopted for $T < T_{order}$ differ in detail from the structure described above for 100% coverage. As before, the ligand chains pack in a hexagonal pattern, but now with the local hexagonal packing rotated by 30° in the plane perpendicular to the average tilt vector (compare

Figures 1d and 1c). At 75-90% coverage, the transformation from the rectangular array of binding sites to the hexagonal packing of the chains ends is achieved by ligands in alternating columns in Figure 1d pointing slightly to alternating sides of the $yz$ plane. This ligand arrangement is characterized by a single peak in $P(\theta_z)$, at larger $\langle\theta_z\rangle$ compared to the ordered state observed at 100% coverage (see the green line in Figure 1g). The two ordered states described thus far are illustrated in the red and green regions of Figure 4a.

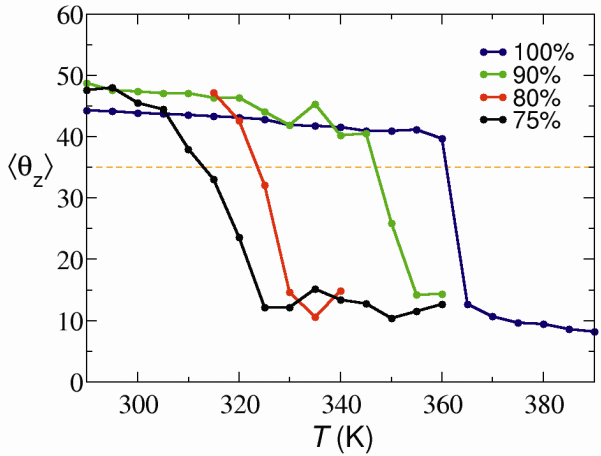

Figure 2. Ligand alignment on the periodic surface as a function of temperature and surface coverage. See Figure 1f and SI for the definition of the tilt angle component $\theta_z$, whose average is plotted here. The 95% confidence interval for these average values is approximately ±6°, and the dashed horizontal line indicates the cutoff used to classify when ligands are strongly ordered ($\langle\theta_z\rangle > 35°$).

**Effect of Changing the Facet Dimensions.** The two ordered states we have described for the periodic CdS(100) surface are distinct from the ordered state observed on CdS(100) facets in our previous study of 4x20 nm CdS nanorods at 100% coverage.[5] On the nanorod, the ligands are strongly tilted not only in the $z$-direction but also in the $x$-direction (towards the edges of the rods). The corresponding peaks in $P(\theta_x)$ around ± 25° are absent on the periodic surface. The transition on the nanorod is also much broader and shifted down in temperature by about 45 K (at equivalent coverage), showing that the facet dimensions strongly affect both the temperature and width of the ordering transition.

To characterize changes in ligand ordering as the nanoscale is approached from above, we consider rod-shaped substrates (like those illustrated in Figure 3) of varying facet width $w$. Their long axis is infinite by virtue of periodic boundary conditions (in the direction $z$ pointing out of the page), and we consider three widths $w$ in the range from ~1 nm to ~10 nm. The macroscopic surfaces characterized above correspond to the limit $w \to \infty$.

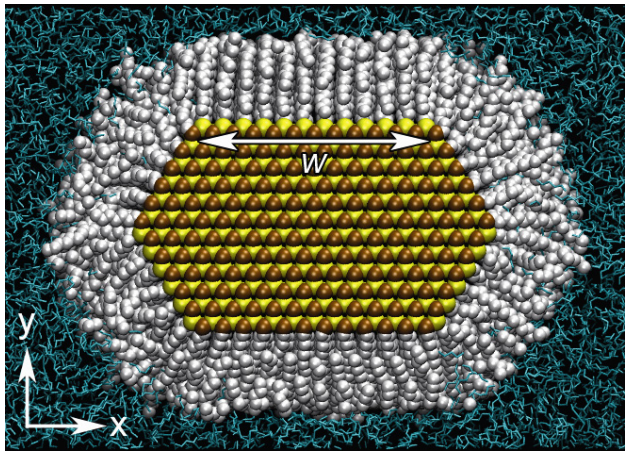

Figure 3. Cross-sectional snapshot of a simulated CdS nanorod in liquid hexane (shown in blue, all other colors as specified in Figure 1) at 290K. This rod spans the simulation cell in the direction $z$ normal to the page, so that the rod is in effect infinitely long. The six side facets correspond to (100) surfaces of CdS. Two of these facets (those perpendicular to the $y$ direction) have a width $w = 4.7$ nm that is larger than that of the other four side facets (whose width is 2.2 nm). We have also examined rods with other values of $w$, specifically $w = 8.0$ nm and $w = 2.2$ nm.

As the facet width $w$ decreases, we find that the structure of the ordered state changes between $w = 2.2$ nm and 4.7 nm, as illustrated in Figure 4a. Specifically, for $w \geq 4.7$ nm the ordered states (labeled I and II in Figure 4) correspond precisely to those formed on a periodic surface at equivalent surface coverage, with little to no tilting of ligands in the $x$-direction, i.e. towards the facet edges. At $w = 2.2$ nm, however, the ordered state (III) features strong tilting of ligands towards the facet edges. This third ordered state, for which symmetry in $\theta_x$ is broken spontaneously, is identical to that formed on 4x20 nm rods[5] (which also have 2.2 nm wide side facets). All three states are characterized by hexagonal packing of the methyl ends of the alkyl ligands, but differ in how the ligands tilt in order to achieve this packing. The states can therefore be distinguished by marked differences in $z$-tilt and $x$-tilt distributions, as shown in Figures 4b and 4c. Near the boundaries between these states, such finite surfaces can adopt multiple ordered states with appreciable probability. Indeed, for $w$ = 4.7 nm at 100% coverage, the top and bottom facets are ordered in states I and II, respectively (see Figure 3).

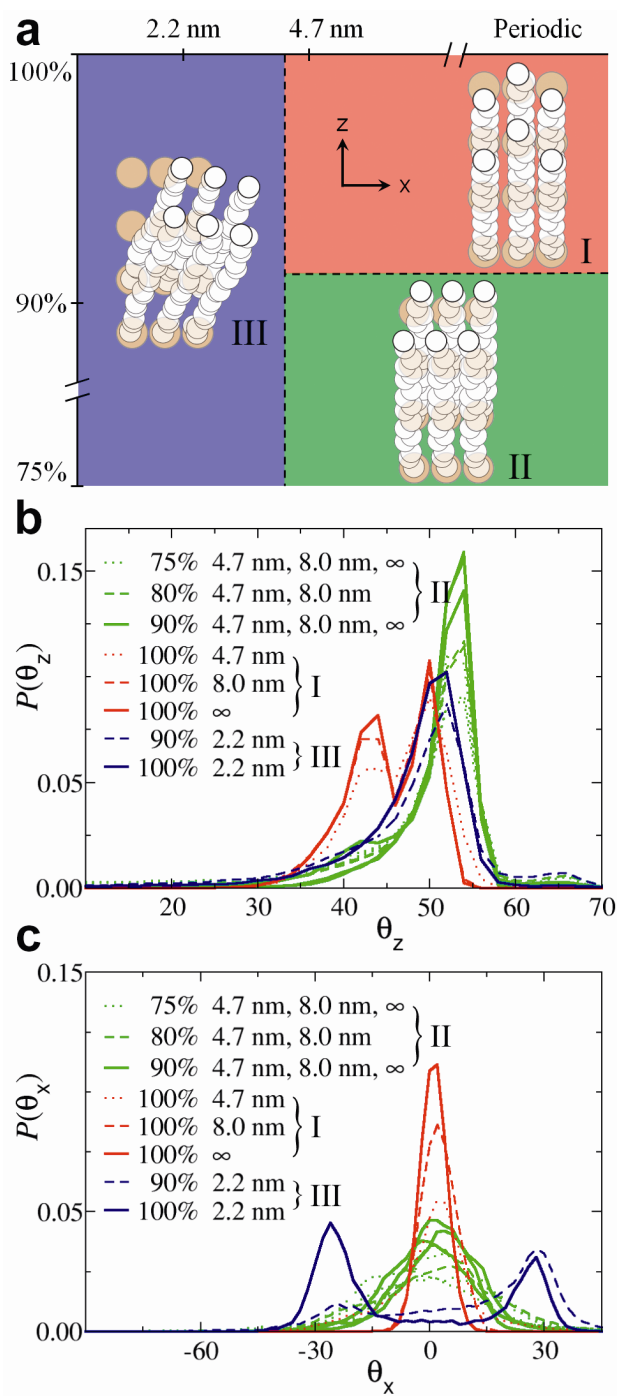

Figure 4. Ordered states of dense ligand shells on CdS(100), and the orientational distribution of molecules within these shells. (a) Schematic illustrating the three different types of packing observed at different values of the facet width and ligand coverage. Phase boundaries (indicated by dashed lines) have been roughly estimated from our results for a handful of values of coverage (100%, 90%, 80%, 75%) and facet width (2.2 nm, 4.7 nm, 8

nm, ∞). The surface Cd atoms are colored brown and the ligands white; other particles are not shown. Note that these configurations are depicted from a different visual perspective than in Figure 1. Probability distributions of tilt angle components - $P(\theta_z)$ and $P(\theta_x)$ - are plotted in (b) and (c) for various geometries and coverage values, colored according to the ordered state that dominates in each case. As in panel (a), red denotes type I order, green denotes type II order, and blue denotes type III order.

Our observation that surface coverage and facet width can change how ligands pack together in the ordered state adds to earlier experiments[9,10] and simulations[18,19] showing that chain length and lattice spacing, respectively, can affect molecular packing in SAMs. In all cases, an interplay between ligand-ligand interactions and constraints imposed by binding to the substrate appear to be responsible for changes, sometimes sudden, in molecular packing.

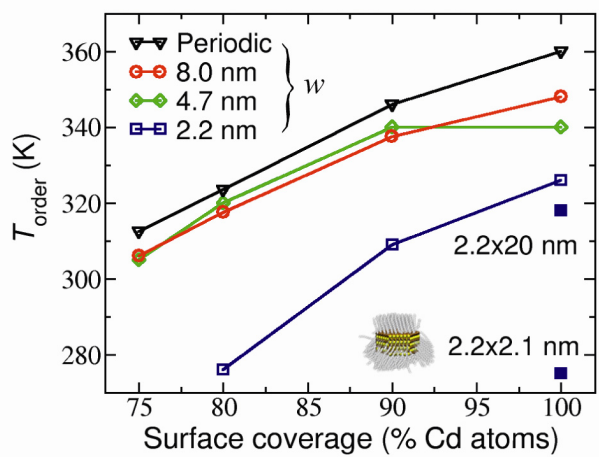

Figure 5. The maximum temperature ($T_{order}$) at which ligands on CdS(100) substrates are strongly ordered ($\langle\theta_z\rangle > 35°$), plotted as a function of the ligand surface coverage. Results are shown for a periodic substrate, and for infinitely long nanorods with several values of the side facet width $w$. We include for comparison results from Ref. 5 for rods of finite length (filled symbols) with the (100) facet dimensions indicated. (The shorter of these "rods" is in fact more disc- than rod-shaped, as shown.)

These changes in ligand structure on nanoscale facets are accompanied by changes in the ordering transition temperature $T_{order}$, as summarized in Figure 5 and indicated by the dashed horizontal lines in Figures 2 and S3.

In particular, these changes are strongly non-linear in their dependence on facet width. For incompletely passivated facets wider than $w = 4.7$ nm, decreasing the facet width effects only modest reduction in $T_{order}$ (by less than 10 K). Further decreasing the width to $w = 2.2$ nm, however, not only changes the symmetry of the ordered state, as discussed above, but also impairs its thermodynamic stability, lowering $T_{order}$ by 30-40 K. The trend of decreasing ordering temperature with decreasing $w$ is less abrupt at 100% coverage, partly because different facets order differently for $w = 4.7$ nm. In contrast to $T_{order}$, the sharpness of the ordering transition does not appear to change substantially with facet width below $w = 8.0$ nm (see Figure S3). Also, in strong contrast to decanethiol monolayers on Au(111), which exhibit a sharp increase in the ordering temperature above 90% coverage,[41] we observe only a small increase in $T_{order}$ going from 90% to 100% coverage.

Both the structure and the stability of these ordered states are thus sensitive to the width of (100) surfaces, which serve as the side facets of CdS nanorods. Figure 5 includes two data points demonstrating sensitivity of ordering to the length of these facets as well. These results, taken from Ref. 5, indicate strong dependence of the ordered state's stability (though not its structure) on rod length. Specifically, at a fixed width of $w = 2.2$ nm, $T_{order}$ decreases by 10 K when the facet length is decreased from infinity to 20 nm and a further 40 K when the length is decreased from 20 nm to 2.1 nm. Finite size effects thus appear to be at least as strong when reducing facet length as compared to facet width. This result is consistent with a recent experimental study of ligand ordering on hexagonal NaYF$_4$ nanoplates,[42] which indicates that facet dimensions terminated by 90° edges can exert stronger finite-size effects than facet dimensions with less acute edges (such as the 60° angle between adjacent CdS(100) facets).

**Solvent Ordering.**

The arrangement of solvent molecules in the vicinity of our ligand-coated surfaces can be quite sensitive to structure in the ligand shell, in some cases exhibiting the strong layering expected for a dense liquid in the vicinity of a flat surface[43]. Here, when ligands are strongly ordered, the $n$-hexane solvent is distinctly layered near the surface, as evidenced by oscillations in average solvent density $\rho_{solv}$ in the direction $y$ perpendicular to the substrate. Such solvent density profiles $\rho_{solv}(y)$ are plotted in Figure 6 for the macroscopic surface at a variety of temperatures and coverages. The magnitude of solvent density oscillations decreases, and the interface eventually broadens, as the ligands lose alignment at lower surface coverage and/or higher temperature. As the passivation density decreases, more solvent molecules penetrate into the ligand shell, and at coverages below ~75% the solvent becomes distinctly layered at the crystal surface itself. There is also a small but clear decrease in the thickness of the ligand shell (by ~0.2 nm), as the boundary between ordered states I and II is crossed upon reducing ligand coverage, due to stronger tilting of ligands towards the substrate in state II.

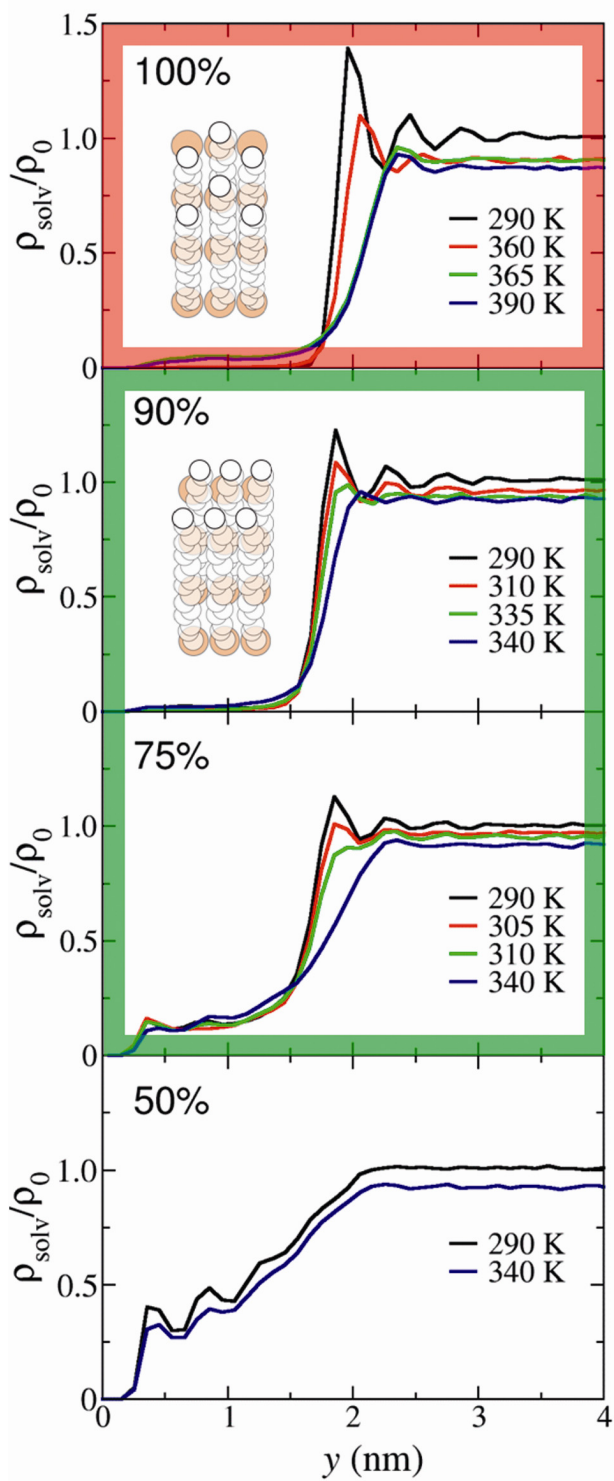

Figure 6. Solvent density profiles normal to a periodic surface. Average densities $\rho_{solv}$, relative to the bulk density $\rho_0$ of *n*-hexane at 290 K, are plotted as a function of distance $y$ from the crystal surface, for several values of ligand coverage (decreasing from top to bottom) and temperature. The colored borders and illustrations indicate the dominant ordered states as in Figure 4a.

These trends with decreasing coverage on periodic substrates are echoed by changes in solvent layering induced by finite facet width. As before, weaker ligand ordering (now on smaller facets) generates less pronounced oscillation in solvent density (see Figure S7). It has previously been noted that solvent ordering between bare nanocrystals weakens as the particle size decreases.[44]

**Surface Forces.**

Our interest in the ligand ordering and solvent layering phenomena discussed above is motivated by their impact on the interaction between passivated nanoscale surfaces. Based on previous results for 4x20 nm nanorods[5] and 2 nm Au particles,[31] we anticipate that disordered ligand shells generally give rise to purely repulsive surface forces in solution,[45] except in the case of very short ligands or of large particles whose cores have large Hamaker constants (such as Au). As ligands order and solvent density layers develop, these interactions are expected to acquire a rich dependence on inter-surface distance, with deep attractive minima near contact. The current study bears out these expectations, showing that changes in ligand ordering lead to pronounced changes in the shape of the PMF.

To characterize ligand- and solvent-mediated interactions between periodic rods like those illustrated in Figure 3, while minimizing computational expense, we constructed systems with opposing surfaces that mimic the junction between nearby nanoparticles. In detail, we explicitly consider two parallel half-rods that are periodically extended in the direction $z$ of their long axes. The simulation cell, including solvent, is also periodically replicated in the lateral $x$ direction (see Figure 7). In order to facilitate calculation of interparticle forces, periodic boundary conditions are *not* imposed in the direction $y$ of particle separation. Instead, we introduce confining walls perpendicular to $y$ via an external potential $U_{wall}(y)$ acting on the solvent molecules and ligands. The potential of mean force (PMF) between the two surfaces is then calculated as detailed in the Methods section.

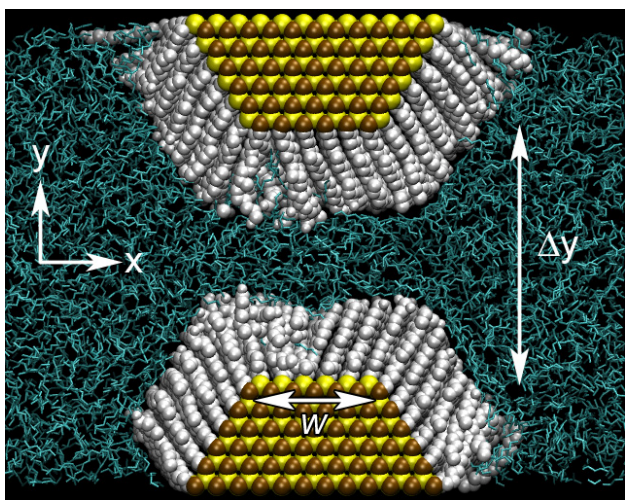

Figure 7. System geometries used to calculate effective interactions between opposing ligand-coated substrates. The simulation cell is periodic in the directions $x$ (left and right) and $z$ (perpendicular to the page). Top and bottom boundaries (parallel to the $xz$-plane) are not periodic, but are instead imposed by a soft repulsive wall. Since CdS nanorods span the simulation cell in the $z$ direction, this setup corresponds to two infinitely long half-rods, in this case with facet width $w$ = 2.2 nm and a rod-rod separation of $\Delta y$ = 5 nm (equilibrated at 290 K). For further details see the SI.

The PMFs plotted in Figure 8 confirm that rod-rod attractions tend to be most potent when ligands are dense and well-ordered, so that solvent density is strongly layered (see Figure 9). Substantial cohesive surface forces persist, however, well below 100% coverage (e.g., between the 4.7 nm facets at 75% coverage) and do not require that ligands on isolated rods are highly aligned. For the case of 2.2 nm facets at 80% coverage, ligands are very weakly aligned at large separations ($y \geq 4.2$ nm) and solvent density is correspondingly featureless in each rod's vicinity; near contact ($y \leq 3.6$ nm), however, ligand alignment is strongly enhanced (see Figure S6), yielding an attraction between rods that is free of oscillations in $y$. Proximity to the order-disorder transition can thus strongly influence surface forces even away from coexistence, much as in the case of drying-induced attractions between nanoscale hydrophobes in liquid water.[46–48] In this case, attractions due to ligand ordering at contact persist at least 10 K above $T_{order}$. Similar contact-induced ligand ordering was recently reported for nanoplates suspended on a liquid surface.[42]

In scale, the PMFs we have computed for different facet widths $w$ at a given surface coverage are roughly proportional to the area of the approaching surfaces. Per unit surface area, well-ordered ligand shells consistently generate a maximum attraction strength of $\phi_{MF}$/area ~2 $kT$/nm$^2$ at $\Delta y$~3.25 nm, followed by decaying oscillations of periodicity ~0.5 nm. The largest and most fully passivated surfaces, specifically the 8.0 nm and 4.7 nm facets at 100% coverage, are the most prominent exceptions. We attribute their smaller scale of attraction to the distinct type I (or mixed type I/II) ordering on these surfaces. The less compact passivation layer of type I ordering also shifts the most probable separation distance from $\Delta y$~3.25 nm to $\Delta y$~3.7 nm. The unusual shape of the PMF for the 4.7 nm facet at 100% coverage may arise from close proximity to coexistence between competing ligand structures, which would suggest a strong sensitivity to changing solution conditions.

We caution that thorough statistical sampling is very difficult to achieve for well-ordered passivation layers on large surfaces, complicating PMF calculations. In these cases, many solvent molecules must move a significant distance in order to evacuate the space between rods as they approach, generating very long relaxation times. We have assessed the uncertainty associated with such slow rearrangements by considering extreme scenarios that, by construction, likely underestimate or overestimate the amount of solvent remaining between rods at equilibrium. An estimated upper bound on the PMF is obtained by sampling at a high approach rate (0.1 nm/ns) that hinders solvent relaxation, generating spurious repulsion on average. To obtain a lower bound on the PMF, we artificially remove solvent from the space between rods (at distances where the ligands are in contact) before re-equilibrating at each separation distance. (See SI for further details.)

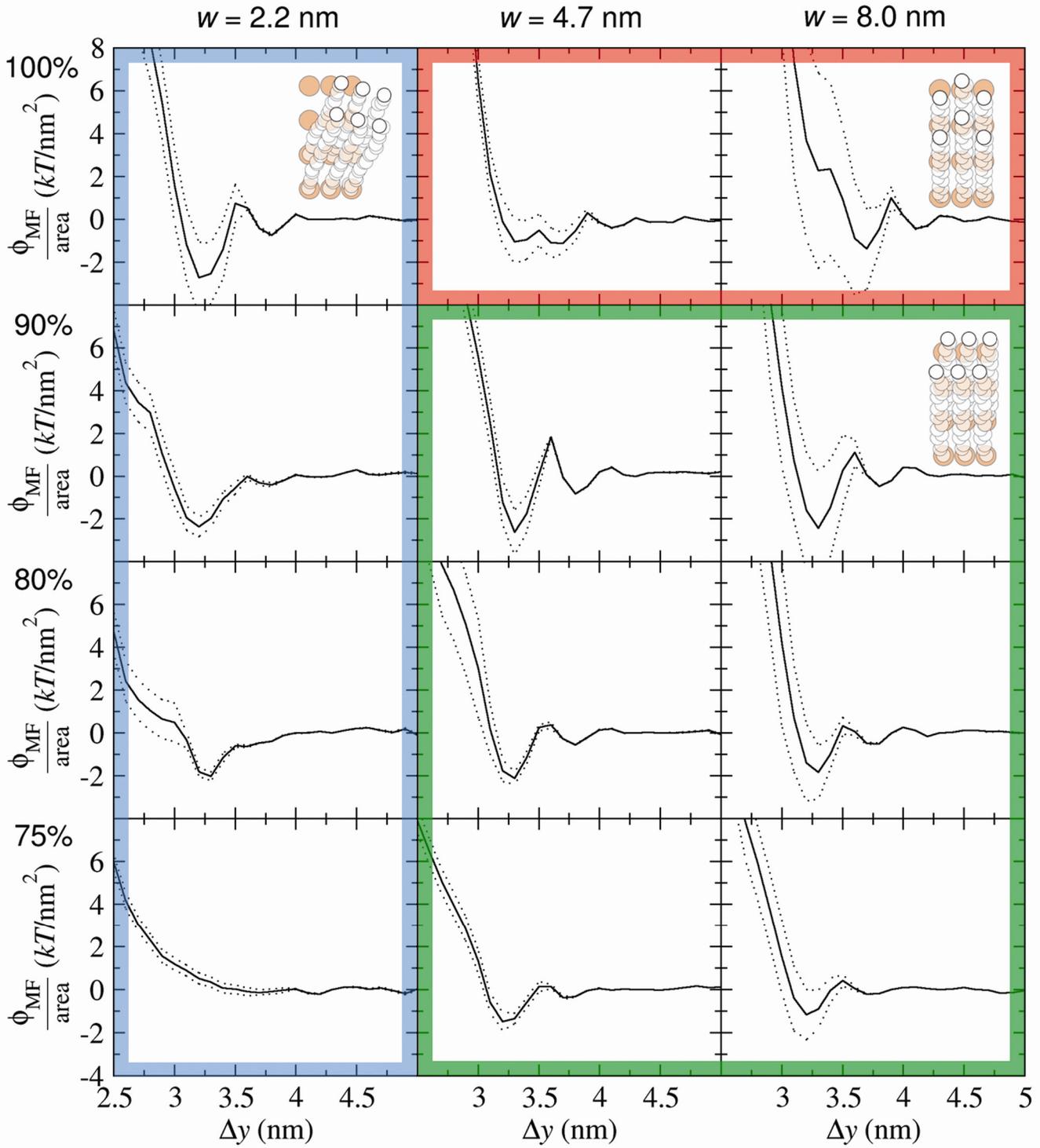

Figure 8. Potentials of mean force ($\phi_{MF}$) between two parallel semi-periodic surfaces (as shown in Figure 7) at 290 K, normalized by the facet area in order to facilitate comparison. Each column of plots shows results for a different facet width $w$, increasing from left to right. Each row shows results for a different surface coverage, decreasing from top to bottom. The dotted lines indicate 95% confidence intervals estimated as described in the text and SI. The colored borders and illustrations indicate the dominant ordered states as in Figure 4a.

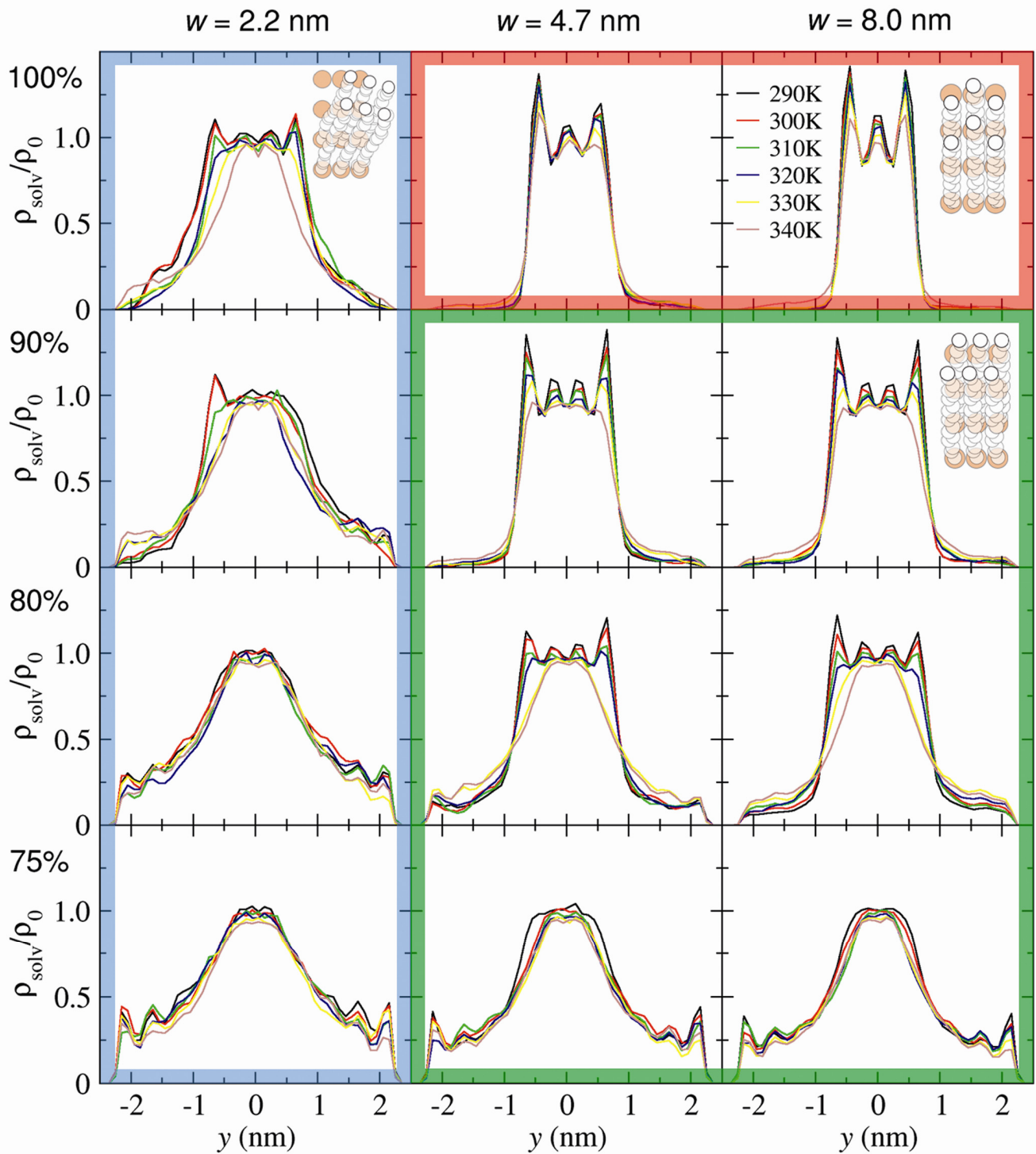

Figure 9. Solvent density profiles in the region between two opposing ligand-coated CdS surfaces (as shown in Figure 7). Average solvent density, relative to the bulk density $\rho_0$ of *n*-hexane at 290 K, is plotted as a function of the coordinate *y*, where 0 indicates the midpoint between the two crystal surfaces. Colors indicate the separation $\Delta y$ between surfaces (see legend at top right). The region used to calculate the solvent density is bounded by the x-coordinates of the two facet edges. The colored borders and illustrations indicate the dominant ordered states as in Figure 4a.

Additional insight into these trends can be gained by dissecting the mean force (as shown in Figure S4) into contributions involving distinct degrees of freedom, e.g., the net force exerted by one substrate's ligands on those of the opposing substrate (see Figure S5). The total force acting between the surfaces is dominated by this ligand-ligand contribution ($F_{LL}$), together with the net force between ligands and solvent molecules ($F_{LS}$). At separations well beyond contact, interactions between ligand shells are attractive ($F_{LL} < 0$), due to dispersion forces that vary slowly with $\Delta y$. At smaller separation, the ligand-ligand force becomes repulsive ($F_{LL} > 0$), despite the fact that the corresponding potential energy remains negative (relative to infinite separation), signaling a potent entropic repulsion when ligand shells restrict one another's conformational freedom. At still closer distances, $F_{LL}$ exhibits in many cases a non-monotonic dependence on $\Delta y$, reflecting competition between deformation of the ligand shells and increasing ligand-ligand contact area. In contrast, $F_{LS}$ is repulsive at all separations considered. Its variation with $\Delta y$ mirrors that of solvent density: $F_{LS}$ is small and nearly constant when solvent is weakly ordered, and it develops substantial oscillations under conditions at which solvent is strongly layered. Damped oscillations in the PMF, with comparable periodicity and decay envelope, result.

Finally, we compare our previous results for discrete 4x20 nm CdS nanorods[5] (which have 2.2 nm wide Cd(100) side facets) with the results presented here for periodic half-rods with $w = 2.2$ nm, both at 100% ligand coverage. These PMFs have similar shape (see Figure S8), but oscillations are stronger for the extended surfaces considered here, with slightly stronger ligand-ligand attraction at close range. These differences could arise from degraded ligand and solvent ordering near the ends of the finite rods. For shorter rods, such finite-size effects can be expected to be even more dramatic, eventually resulting in the loss of any thermally significant attraction between nanoparticles.

## DISCUSSION

This work has shown that alkane ligand monolayers on nanoscale surfaces can adopt multiple distinct ordered states, depending on surface coverage and facet dimensions. The anisotropic CdS(100) substrates we consider give rise to ordering scenarios that differ substantially from those reported for the more symmetric Au(111) surface. For example, ligand alignment emerges more sharply on CdS due to the reduced symmetry of the Wurzite (100) surface.

Implications of these findings for real systems are complicated by the unavoidably approximate nature of the model we have studied. For small quantum dots, different united-atom potentials can give very different predictions of ligand structure at a given temperature.[49] Viewed over a broad range of temperature, however, the general phase behaviors of these different models are consistent with one another, and also consistent with more detailed all-atom simulations. As in molecular simulations of many systems,[50,51] the variety and structures of equilibrium states we have described are likely to be more robust predictions than are their relative stabilities. Among all our results, the ordering temperatures we have predicted are thus subject to the greatest uncertainty. We nonetheless expect these transition temperatures to be reasonably accurate, given the similarity (in form and scale) of the force field studied here to that of Paul *et al.*,[52] which has been vetted against all-atom calculations.[49] Indeed, for periodic surfaces at 100% coverage in *n*-hexane, the value of $T_{\text{order}}$ we have determined lies within 5 K of that obtained with the force field of Ref. 52. In light of our conclusion that ordering temperatures can be very sensitive to particle size and shape, as well as ligand coverage, a direct quantitative comparison with experiment would require a closer and more systematic control over these features than has yet been reported. For example, experimental samples are not monodisperse, their ligand coverage can vary strongly with handling,[39,40] and surfaces can be enriched with metal atoms.[25] Like most computer simulations of nanoparticles, our model further assumes that surface atoms and ligand attachment sites are immobile. Recent experiments that reveal significant mobility of surface atoms on Au nanorods[53] suggest that this common approximation deserves careful scrutiny.

Our work also shows that, as a consequence of this rich phase behavior, the effective interaction between opposing surfaces is a sensitive function of geometry and solution conditions. For example, the effects of changing temperature can be much more complex than modifying the Boltzmann weight of a surface attraction with unvarying strength. By changing the nature or degree of ligand ordering, control variables such as temperature can instead modulate surface forces sharply and non-monotonically, with important consequences for the solution stability and assembly of ligand-coated nanoparticles. This sensitivity could contribute to the difficulty of achieving highly reproducible self-assembly in experiments, which have in many cases not exerted control over ligand coverage.

The sharpness of the ordering transition on the anisotropic CdS(100) surface could present another problem for reproducible assembly, in that the strength of attraction changes dramatically over a small temperature interval and so may be difficult to fine-tune. Our results indicate that decreasing the surface coverage does not help much to broaden the transition. The broader ordering transition observed on the more isotropic Au(111) surface indicates that fine-tuning the interaction between nanoparticles might be easier when ligand binding sites are distributed more isotropically on their facets.

The ordered ligand states formed on CdS(100) differ distinctly from those on Au(111). Our most general finding – that ligand ordering can induce attraction between particles in solution – is nonetheless consistent with recent experimental results for Au nanoparticles coated with $C_{12}$-$C_{18}$ alkanethiol ligands in *n*-heptane solvent.[54,55] In that study reversible aggregation was observed upon cooling and heating,[55] while an optimal temperature region was found for the formation of ordered crystalline aggregates when the aggregation was driven by the addition of a poor polar solvent.[54] Such optimum assembly often marks the boundary of thermodynamic stability for the assembled state.[56] If nanoparticle aggregation were driven by ligand ordering, we would thus expect conditions of optimal assembly to track equilibrium coexistence curves for the respective pure alkanes and alkanethiol ligands on dry nanoparticles. Indeed, the observed variation of nanoparticle aggregation temperature with chain length closely reflects the ligands' equilibrium phase behavior.[55]

Finally, the complex ligand ordering we have characterized on CdS(100) may strongly influence functional properties of nanoparticles, including their use in lubrication, biological sensing, and optoelectronics. The exchange of charge carriers with other particles or electrodes, for instance, can depend sensitively on particle separation, which we have shown to vary with structure of the ligand layer. In cases where charge transport mechanisms depend on surface defects and tunneling barriers, the role of ligand ordering could be even more important.[25] As another example, our results indicate that it might be possible to reversibly and dynamically tune the frictional properties of nanoparticles, which are being explored as lubricants,[57,58] by altering the structure of their ligand shell.

## CONCLUSIONS

Overall, our results show that there are strong similarities, as well as some significant differences, between the behavior of non-polar ligands on macroscopic substrates in non-polar solvent and that on nanoscale surfaces. In both cases, the ligands align strongly with one another at sufficiently low temperature to form a self-assembled monolayer (provided that the surface coverage is sufficiently high), and in both cases alignment results in a substantial increase in attraction between ligand layers on opposing substrates. Substantial finite size effects emerge in our studies only on facets less than 4.7 nm in width. For such facets, highly aligned ligand states are less thermodynamically stable, and ordering is less cooperative, as is often the case for phase transitions in small systems. More surprisingly, the structure and symmetry of ordered states can vary with facet dimension. The edge-ward tilting of ligands we have observed on narrow facets significantly influences PMFs between opposing surfaces. Diminished solvent layering near nanoscale facets also impacts the strength and distance dependence of surface interactions.

Our results thus add to a growing body of evidence that ligand-mediated interactions can exert a particularly strong influence on the assembly of faceted nanoparticles.[40,59–61] The strongly temperature-dependent mode of interaction we have examined suggests new ways to control assembly by modulating ligand structure, e.g., reversibly activating strong face-to-face attraction between ordered ligand layers through modest changes in temperature. Our finding that transition temperatures are highly sensitive to surface coverage further suggests the possibility of selectively introducing attraction between particular sets of facets based on differences in their ligand coverages. This approach could enable straightforward realization of patchy particles that organize into non-compact superlattices (such as the open square-hexagon network described in Ref. 62), a challenging and important goal for synthetic systems.

Assessing the quantitative accuracy of our computational findings in many cases awaits technical advances in the laboratory. Some of the trends and qualitative observations we have described, however, should be accessible to current experimental techniques. For example, it should be possible to distinguish between different ligand ordered states on a planar CdS(100) surface, or in films of dry CdS nanoparticles, and to study shifts in the order-disorder transition upon changing temperature or surface coverage. Contributions from different facets could be discriminated by varying the aspect ratio of rod-shaped particles, or by studying nanoparticles lying at the surface of a thin liquid film.[42] In the latter case, surface energies should cause CdS nanorods to align in the plane of the film, thus allowing ligand ordering on the (100) side facets to be probed by a surface-sensitive spectroscopy such as SFG. This type of measurement could be combined with a Langmuir trough force measurement to study the connection between the ligand order on the nanoparticles and the particle-particle interaction.

**METHODS**

In this paper we have studied several models, illustrated in Figures 1, 3 and 7. These models consist of wurzite CdS crystals with exposed (100) facets passivated by octadecyl ligands in an explicit *n*-hexane solvent. The (100) surface is the dominant side facet observed in HRTEM images of CdS and CdSe nanorods.[37] The ligands were bound to surface Cd atoms, with the coverage ranging from 50% to 100%. The pattern of passivation was random at 90%, but periodic at lower values to ensure even surface coverage (see Figure S1). For example, at 50% coverage we only passivated every second Cd atom in the *x*-direction, as this lattice dimension (0.412 nm) is substantially smaller than the Cd spacing in the perpendicular *z*-direction (0.675 nm). The periodic *z*-dimension was 8.1004 nm in all models, and the *x*-dimension was 8.24293 nm for the periodic planar surfaces. Further details are provided in the SI.

The ligand and solvent molecules were modeled using a united-atom potential, which represents each $CH_x$ group with a single particle. Interactions between these coarse-grained particles include volume exclusion and dispersion as described by the Lennard-Jones (LJ) potential and, within each molecule, bond stretching, bond bending, and dihedral torsion terms. The $CH_x$ groups interact with CdS atoms *via* a Lennard-Jones potential previously developed for Au nanoparticles,[16] adapted to reflect the Hamaker constant and density of CdS rather than Au. Further details of these potentials are provided in Ref. 5.

Molecular dynamics (MD) simulations on systems of up to 42,000 particles were performed using LAMMPS,[63] at temperatures ranging from 270 K to 390 K. Because the CdS crystalline lattice is quite rigid, and the related CdSe lattice has been found to undergo minimal surface reconstruction when passivated,[64] we held the positions of Cd and S atoms fixed in all simulations, thus saving considerable computational expense. Below the standard boiling point of *n*-hexane (342 K), the bulk solvent density was tuned to the experimental density of pure *n*-hexane at 1 atm, while above this it was tuned to the density of liquid *n*-hexane at the corresponding state point on its liquid-vapor coexistence curve.[65] Systems were equilibrated at fixed volume and constant temperature, maintained with a Nosé-Hoover thermostat, for 1-3 ns, and subsequent production runs (also at fixed volume and temperature) were 0.5-2 ns in duration.

For the semi-periodic setups illustrated in Figure 7, we used constrained MD to calculate the potential of mean force (PMF) between the two parallel surfaces, as a function of their normal separation $\Delta y$. The mean force between the surfaces ($F_{mean}$) is given by the average force in the direction of their connecting line:

$$F_{mean}(\Delta y) = \frac{1}{2}\langle (\mathbf{F_2} - \mathbf{F_1}) \cdot \hat{\mathbf{y}} \rangle_{NVT} \qquad (1)$$

where $\mathbf{F_2}$ and $\mathbf{F_1}$ are the total forces acting on the two surfaces, $\hat{\mathbf{y}}$ is the unit vector normal to both surfaces, and angular brackets denote an average in the canonical ensemble. The PMF is then given by:

$$\phi_{MF}(\Delta y) = \int_{\Delta y}^{\infty} [F_{mean}(s) - F_{mean}(\infty)]\, ds \qquad (2)$$

where $F_{mean}(\infty)$ is the force between the surfaces before they start to interact. This background force arises from the presence of solvent on only one side of each surface, which is necessary to calculate the PMF using the setup shown in Figure 7, which employs fixed boundaries at the top and bottom of the simulation cell. In practice, $F_{mean}(\infty)$ was calculated by averaging $F_{mean}(\Delta y)$ over nine independent separations at distances beyond which there was significant interaction between the surfaces. The interaction between the CdS cores is negligible at all separations we have considered.

**Acknowledgement.** This work was supported by generous grants of computer time from the National Computational Infrastructure facility (which is supported by the Australian Government) and the National Energy Research Scientific Computing Center (which was supported by the Director, Office of Science, Office of Basic Energy Sciences, Materials Sciences, and Engineering Division, of the U.S. Department of Energy under contract no. DE-AC02-05CH11231). A.W. acknowledges financial support from the Australian Research Council in the form of a Future Fellowship (FT140101061).

**Supporting Information Available.** Additional model and simulation details, and additional data on ligand alignment and the interaction between CdS(100) surfaces in hexane. This material is available free of charge *via* the Internet at http://pubs.acs.org.

# REFERENCES AND NOTES

(1) Glotzer, S. C.; Solomon, M. J. Anisotropy of Building Blocks and Their Assembly into Complex Structures. *Nat. Mater.* **2007**, *6*, 557–562.

(2) Peng, X.; Manna, L.; Yang, W.; Wickham, J.; Scher, E.; Kadavanich, A.; Alivisatos, A. P. Shape Control of CdSe Nanocrystals. *Nature* **2000**, *404*, 59–61.

(3) Wuister, S. F.; van Houselt, A.; de Mello Donegá, C.; Vanmaekelbergh, D.; Meijerink, A. Temperature Antiquenching of the Luminescence from Capped CdSe Quantum Dots. *Angew. Chemie* **2004**, *43*, 3029–3033.

(4) Silvera Batista, C. A.; Larson, R. G.; Kotov, N. A. Nonadditivity of Nanoparticle Interactions. *Science* **2015**, *350*, 1242477.

(5) Widmer-Cooper, A.; Geissler, P. Orientational Ordering of Passivating Ligands on CdS Nanorods in Solution Generates Strong Rod-Rod Interactions. *Nano Lett.* **2014**, *14*, 57–65.

(6) Ulman, A. Formation and Structure of Self-Assembled Monolayers. *Chem. Rev.* **1996**, *96*, 1533–1554.

(7) Love, J. C.; Estroff, L. A.; Kriebel, J. K.; Nuzzo, R. G.; Whitesides, G. M. Self-Assembled Monolayers of Thiolates on Metals as a Form of Nanotechnology. *Chem. Rev.* **2005**, *105*, 1103–1169.

(8) Schreiber, F. Self-Assembled Monolayers: From " Simple " Model Systems to Biofunctionalized Interfaces. *J. Phys. Condens. Matter* **2004**, *16*, R881–R900.

(9) Fenter, P.; Eisenberger, P.; Liang, K. Chain-Length Dependence of the Structures and Phases of $CH_3(CH_2)_{n-1}SH$ Self-Assembled on Au (111). *Phys. Rev. Lett.* **1993**, *70*, 2447–2450.

(10) Fenter, P.; Eberhardt, A.; Liang, K. S.; Eisenberger, P. Epitaxy and Chainlength Dependent Strain in Self-Assembled Monolayers. *J. Chem. Phys.* **1997**, *106*, 1600–1608.

(11) Ninham, B. W.; Christenson, H. K. Interaction of Hydrocarbon Monolayer Surfaces across N-Alkanes: A Steric Repulsion. *J. Chem. Phys.* **1989**, *90*, 5801–5805.

(12) Gee, M.; Israelachvili, J. N. Interactions of Surfactant Monolayers across Hydrocarbon Liquids. *J. Chem. Soc., Faraday Trans.* **1990**, *86*, 4049–4058.

(13) Israelachvili, J. N. Repulsive "Steric" or "Overlap" Forces between Polymer-Covered Surfaces. In *Intermolecular and Surface Forces (3rd Edition)*; Academic Press, 2011; pp. 387–392.

(14) Pertsin, A. J.; Hayashi, T.; Grunze, M. Grand Canonical Monte Carlo Simulations of the Hydration Interaction between Oligo(ethylene Glycol)-Terminated Alkanethiol Self-Assembled Monolayers. *J. Phys. Chem. B* **2002**, *106*, 12274–12281.

(15) Hautman, J.; Klein, M. L. Molecular Dynamics Simulation of the Effects of Temperature on a Dense Monolayer of Longchain Molecules. *J. Chem. Phys.* **1990**, *93*, 7483–7492.

(16) Pool, R.; Schapotschnikow, P.; Vlugt, T. J. H. Solvent Effects in the Adsorption of Alkyl Thiols on Gold Structures: A Molecular Simulation Study. *J. Phys. Chem. C* **2007**, *111*, 10201–10212.

(17) Mar, W.; Klein, M. L. Molecular Dynamics Study of the Self-Assembled Monolayer Composed of S(CH2)14CH3 Molecules Using an All-Atoms Model. *Langmuir* **1994**, *10*, 188–196.

(18) Schmid, F.; Stadler, C.; Düchs, D. Computer Simulations of Self-Assembled Monolayers. *J. Phys. Condens. Matter* **2001**, *13*, 8653–8659.

(19) Vemparala, S.; Karki, B. B.; Kalia, R. K.; Nakano, A.; Vashishta, P. Large-Scale Molecular Dynamics Simulations of Alkanethiol Self-Assembled Monolayers. *J. Chem. Phys.* **2004**, *121*, 4323–4330.

(20) Hostetler, M. J.; Wingate, J. E.; Zhong, C.; Harris, J. E.; Vachet, R. W.; Clark, M. R.; Londono, J. D.; Green, S. J.; Stokes, J. J.; Wignall, G. D.; Glish, G. L.; Porter, M. D.; Evans, N. D.; Murray, R. W. Alkanethiolate Gold Cluster Molecules with Core Diameters from 1.5 to 5.2 Nm: Core and Monolayer Properties as a Function of Core Size. *Langmuir* **1998**, *14*, 17–30.

(21) Badia, A.; Cuccia, L.; Demers, L.; Morin, F.; Lennox, R. B. Structure and Dynamics in Alkanethiolate Monolayers Self-Assembled on Gold Nanoparticles: A DSC, FT-IR, and Deuterium NMR Study. *J. Am. Chem. Soc.* **1997**, *119*, 2682–2692.

(22) Luedtke, W. D.; Landman, U. Structure, Dynamics, and Thermodynamics of Passivated Gold Nanocrystallites and Their Assemblies. *J. Phys. Chem.* **1996**, *100*, 13323–13329.

(23) Luedtke, W. D.; Landman, U. Structure and Thermodynamics of Self-Assembled Monolayers on Gold Nanocrystallites. *J. Phys. Chem. B* **1998**, *5647*, 6566–6572.

(24) Ghorai, P. K.; Glotzer, S. C. Molecular Dynamics Simulation Study of Self-Assembled Monolayers of Alkanethiol Surfactants on Spherical Gold Nanoparticles. *J. Phys. Chem. C* **2007**, *111*, 15857–15862.

(25) Frederick, M. T.; Achtyl, J. L.; Knowles, K. E.; Weiss, E. a.; Geiger, F. M. Surface-Amplified Ligand Disorder in CdSe Quantum Dots Determined by Electron and Coherent Vibrational Spectroscopies. *J. Am. Chem. Soc.* **2011**, *133*, 7476–7481.

(26) Landman, U.; Luedtke, W. D. Small Is Different: Energetic, Structural, Thermal, and Mechanical Properties of Passivated Nanocluster Assemblies. *Faraday Discuss.* **2004**, *125*, 1–22.

(27) Wang, Z.; Schliehe, C.; Bian, K.; Dale, D.; Bassett, W. A.; Hanrath, T.; Klinke, C.; Weller, H. Correlating Superlattice Polymorphs to Internanoparticle Distance, Packing Density, and Surface Lattice in Assemblies of PbS Nanoparticles. *Nano Lett.* **2013**, *13*, 1303–1311.

(28) Templeton, A. C.; Hostetler, M. J.; Kraft, C. T.; Murray, R. W.; Hill, C.; Carolina, N. Reactivity of Monolayer-Protected Gold Cluster Molecules: Steric Effects. *J. Am. Chem. Soc.* **1998**, *120*, 1906–1911.

(29) Wuelfing, W. P.; Templeton, A. C.; Hicks, J. F.; Murray, R. W. Taylor Dispersion Measurements of Monolayer Protected Clusters: A Physicochemical Determination of Nanoparticle Size. *Anal. Chem.* **1999**, *71*, 4069–4074.

(30) Lane, J. M. D.; Grest, G. S. Spontaneous Asymmetry of Coated Spherical Nanoparticles in Solution and at Liquid-Vapor Interfaces. *Phys. Rev. Lett.* **2010**, *104*, 235501.

(31) Schapotschnikow, P.; Pool, R.; Vlugt, T. J. H. T. Molecular Simulations of Interacting Nanocrystals. *Nano Lett.* **2008**, *8*, 2930–2934.

(32) Lane, J. M. D.; Ismail, A. E.; Chandross, M.; Lorenz, C. D.; Grest, G. S. Forces between Functionalized Silica Nanoparticles in Solution. *Phys. Rev. E* **2009**, *79*, 050501(R).

(33) Yang, Z.; Yang, X.; Xu, Z.; Yang, N. Molecular Simulations of Structures and Solvation Free Energies of Passivated Gold Nanoparticles in Supercritical CO 2. *J. Phys. Chem.* **2010**, *133*, 094702.

(34) Woodward, J.; Ulman, A.; Schwartz, D. Self-Assembled Monolayer Growth of Octadecylphosphonic Acid on Mica. *Langmuir* **1996**, *7463*, 3626–3629.

(35) Anderson, N. C.; Hendricks, M. P.; Choi, J. J.; Owen, J. S. Ligand Exchange and the Stoichiometry of Metal Chalcogenide Nanocrystals: Spectroscopic Observation of Facile Metal-Carboxylate Displacement and Binding. *J. Am. Chem. Soc.* **2013**, *135*, 18536–18548.

(36) Knittel, F.; Gravel, E.; Cassette, E.; Pons, T.; Pillon, F.; Dubertret, B.; Doris, E. On the Characterization of the Surface Chemistry of Quantum Dots. *Nano Lett.* **2013**, *13*, 5075–5078.

(37) Rosenthal, S. J.; McBride, J.; Pennycook, S. J.; Feldman, L. C. Synthesis, Surface Studies, Composition and Structural Characterization of CdSe, Core/Shell, and Biologically Active Nanocrystals. *Surf. Sci. Rep.* **2007**, *62*, 111–157.

(38) Morris-Cohen, A. J.; Malicki, M.; Peterson, M. D.; Slavin, J. W. J.; Weiss, E. a. Chemical, Structural, and Quantitative Analysis of the Ligand Shells of Colloidal Quantum Dots. *Chem. Mater.* **2013**, *25*, 1155–1165.

(39) Moreels, I.; Fritzinger, B.; Martins, J. C.; Hens, Z. Surface Chemistry of Colloidal PbSe Nanocrystals. *J. Am. Chem. Soc.* **2008**, *130*, 15081–15086.

(40) Choi, J. J.; Bealing, C. R.; Bian, K.; Hughes, K. J.; Zhang, W.; Smilgies, D. M.; Hennig, R. G.; Engstrom, J. R.; Hanrath, T. Controlling Nanocrystal Superlattice Symmetry and Shape-Anisotropic Interactions through Variable Ligand Surface Coverage. *J. Am. Chem. Soc.* **2011**, *133*, 3131–3138.

(41) Schreiber, F.; Eberhardt, a.; Leung, T.; Schwartz, P.; Wetterer, S.; Lavrich, D.; Berman, L.; Fenter, P.; Eisenberger, P.; Scoles, G. Adsorption Mechanisms, Structures, and Growth Regimes of an Archetypal Self-Assembling System: Decanethiol on Au(111). *Phys. Rev. B* **1998**, *57*, 12476–12481.

(42) Zhang, H.; Li, F.; Xiao, Q.; Lin, H. Conformation of Capping Ligands on Nanoplates: Facet-Edge-Induced Disorder and Self-Assembly-Related Ordering Revealed by Sum Frequency Generation Spectroscopy. *J. Phys. Chem. Lett.* **2015**, *6*, 2170–2176.

(43) Israelachvili, J. N. Molecular Ordering at Surfaces, Interfaces, and in Thin Film. In *Intermolecular and Surface Forces (3rd Edition)*; Academic Press, 2011; pp. 342–343.

(44) Qin, Y.; Fichthorn, K. A. Molecular Dynamics Simulation of the Forces between Colloidal Nanoparticles in N-Decane Solvent. *J. Chem. Phys.* **2007**, *127*, 144911.

(45) In vacuum we have previously found that the interaction is highly attractive regardless of the ligand order.[5] We expect similar behavior for the lower surface coverage scenarios explored in this paper.

(46) Lum, K.; Chandler, D.; Weeks, J. D. Hydrophobicity at Small and Large Length Scales. *J. Phys. Chem. B* **1999**, *103*, 4570–4577.

(47) Ten Wolde, P. R.; Chandler, D. Drying-Induced Hydrophobic Polymer Collapse. *Proc. Natl. Acad. Sci. USA* **2002**, *99*, 6539–6543.

(48) Miller, T. F.; Vanden-Eijnden, E.; Chandler, D. Solvent Coarse-Graining and the String Method Applied to the Hydrophobic Collapse of a Hydrated Chain. *Proc. Natl. Acad. Sci. USA* **2007**, *104*, 14559–14564.

(49) Kaushik, A. P.; Clancy, P. Explicit All-Atom Modeling of Realistically Sized Ligand-Capped Nanocrystals. *J. Chem. Phys.* **2012**, *136*, 114702.

(50) García Fernández, R.; Abascal, J. L. F.; Vega, C. The Melting Point of Ice Ih for Common Water Models Calculated from Direct Coexistence of the Solid-Liquid Interface. *J. Chem. Phys.* **2006**, *124*, 1–11.

(51) Aragones, J. L.; Sanz, E.; Valeriani, C.; Vega, C. Calculation of the Melting Point of Alkali Halides by Means of Computer Simulations. *J. Chem. Phys.* **2012**, *137*, 104507.

(52) Paul, W.; Yoon, D. Y.; Smith, G. D. An Optimized United Atom Model for Simulations of Polymethylene Melts. *J. Chem. Phys.* **1995**, *103*, 1702–1709.

(53) Katz-Boon, H.; Walsh, M.; Dwyer, C.; Mulvaney, P.; Funston, A. M.; Etheridge, J. Stability of Crystal Facets in Gold Nanorods. *Nano Lett.* **2015**, *15*, 1635–1641.

(54) Geyer, T.; Born, P.; Kraus, T. Switching Between Crystallization and Amorphous Agglomeration of Alkyl Thiol-Coated Gold Nanoparticles. *Phys. Rev. Lett.* **2012**, *109*, 128302.

(55) Born, P.; Kraus, T. Ligand-Dominated Temperature Dependence of Agglomeration Kinetics and Morphology in Alkyl-Thiol-Coated Gold Nanoparticles. *Phys. Rev. E* **2013**, *87*, 062313.

(56) Whitelam, S.; Feng, E.; Hagan, M.; Geissler, P. The Role of Collective Motion in Examples of Coarsening and Self-Assembly. *Soft Matter* **2009**, *5*, 1251–1262.

(57) Min, Y.; Akbulut, M.; Prud'homme, R. K.; Golan, Y.; Israelachvili, J. Frictional Properties of Surfactant-Coated Rod-Shaped Nanoparticles in Dry and Humid Dodecane. *J. Phys. Chem. B* **2008**, *112*, 14395–14401.

(58) Jones, R. L.; Pearsall, N. C.; Batteas, J. D. Disorder in Alkylsilane Monolayers Assembled on Surfaces with Nanoscopic Curvature. *J. Phys. Chem. C* **2009**, *113*, 4507–4514.

(59) Jones, M. R.; Macfarlane, R. J.; Prigodich, A. E.; Patel, P. C.; Mirkin, C. A. Nanoparticle Shape Anisotropy Dictates the Collective Behavior of Surface-Bound Ligands. *Nature* **2011**, *133*, 18865–18869.

(60) Henzie, J.; Grünwald, M.; Widmer-Cooper, A.; Geissler, P. L.; Yang, P. Self-Assembly of Uniform Polyhedral Silver Nanocrystals into Densest Packings and Exotic Superlattices. *Nat. Mater.* **2012**, *11*, 131–137.

(61) Ye, X.; Chen, J.; Engel, M.; Millan, J. A.; Li, W.; Qi, L.; Xing, G.; Collins, J. E.; Kagan, C. R.; Li, J.; Glotzer, S. C.; Murray, C. B. Competition of Shape and Interaction Patchiness for Self-Assembling Nanoplates. *Nat. Chem.* **2013**, *5*, 466–473.

(62) Millan, J. A.; Ortiz, D.; Van Anders, G.; Glotzer, S. C. Self-Assembly of Archimedean Tilings with Enthalpically and Entropically Patchy Polygons. *ACS Nano* **2014**, *8*, 2918–2928.

(63) Plimpton, S. J. Fast Parallel Algorithms for Short-Range Molecular Dynamics. *J. Comp. Phys.* **1995**, *117*, 1–19 and http://lammps.sandia.gov/.

(64) Manna, L.; Wang, L. W.; Cingolani, R.; Alivisatos, A. P. First-Principles Modeling of Unpassivated and Surfactant-Passivated Bulk Facets of Wurtzite CdSe: A Model System for Studying the Anisotropic Growth of CdSe Nanocrystals. *J. Phys. Chem. B* **2005**, *109*, 6183–6192.

(65) Beg, S. A.; Tukur, N. M.; Al-Harbi, D. K.; Hamad, E. Z. Saturated Liquid Densities of Benzene, Cyclohexane, and Hexane from 298.15 to 473.15 K. *J. Chem. Eng. Data* **1993**, *38*, 461–464.

**Table of contents graphic**

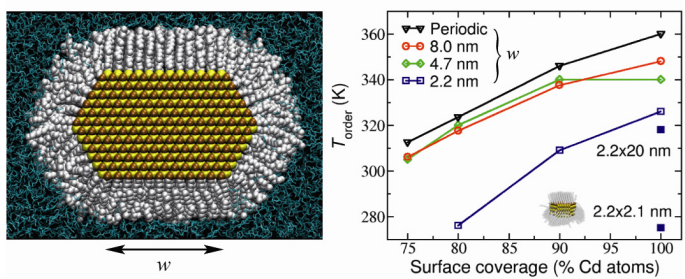